\documentclass[preprint, pre,superscriptaddress, letterpaper, fleqn, floatfix, showpacs ]{revtex4}

\usepackage{euscript, units, amssymb,amsfonts,amsmath, graphics, graphicx, dcolumn, fancyhdr}

\usepackage{times}
\usepackage{hyperref}

\usepackage{graphics, graphicx}

\begin{document}

\title{Wave spectra of 2D Yukawa solids and liquids in the presence of a magnetic field}
\author{Lu-Jing Hou}\email{ljhouwang@gmail.com}
\affiliation{IEAP, Christian-Albrechts Universit\"{a}t zu Kiel, D-24098 Kiel, Germany}

\author{P. K. Shukla}
\affiliation{Institut f\"{u}r Theoretische Physik IV, Ruhr-Universit\"{a}t Bochum, D-44780, Germany}
\author{Alexander Piel}
\affiliation{IEAP, Christian-Albrechts Universit\"{a}t zu Kiel, D-24098 Kiel, Germany}
\author{Z.\ L.\ Mi\v{s}kovi\'{c}}
\affiliation{Department of Applied Mathematics, University of Waterloo, Waterloo, Ontario, Canada N2L 3G1}

\date{\today}

\begin{abstract}
Thermally excited phonon spectra of 2D Yukawa solids and liquids in the presence of an external magnetic field are studied using
computer simulations. Special attention is paid to the variation of wave spectra in terms of several key parameters, such as the
strength of coupling, the screening parameter, and the intensity of the magnetic field. In addition, comparisons are made with
several analytical theories, including random-phase approximation, quasi-localized charge approximation, and harmonic
approximation, and the validity of those theories is discussed in the present context.
\end{abstract}
 \maketitle



\section{Introduction}
There has been a great deal of interest in studying collective dynamics of two-dimensional (2D) strongly coupled Coulomb systems
(SCCSs) in the presence of an external magnetic field
\cite{Chiu1974,Fukuyama1975,Bonsall1977,Cote1991,Golden1993,Ranganathan2002,Ranganathan2005}. Early studies were mainly
concerned with the 2D electron gas and/or liquid systems, and were largely motivated by their potential applications in the
semiconductor industry, e.g., for metal-oxide-semiconductor (MOS) structures \cite{Chiu1974,Fukuyama1975,Bonsall1977,Cote1991}.
Recently, interests have turned to a slightly different system, i.e., the so-called strongly coupled complex (dusty) plasmas
(SCDPs) \cite{Shukla2002,Shukla2009}.

Complex (dusty) plasma is a suspension of micron-sized charged dust particles in a weakly ionized plasma with electrons, ions,
and neutral atoms or molecules \cite{Shukla2002,Shukla2009}. Therein, dust particles acquire a few thousand electron charges by
absorbing surrounding electrons and ions, and consequently interact with each other via a dynamically screened Coulomb potential
\cite{Lampe2000} while undergoing Brownian motion due primarily to frequent collisions with the neutral molecules. When the
interaction potential energy between charged dust particles significantly exceeds their kinetic energy, they become strongly
coupled and can form ordered structures comprising liquid and solid states. Since the motion of charged dust particles in
complex (dusty) plasmas can be directly observed in real time by using a video camera, such systems have been generally regarded
as a promising model system to study many phenomena occurring in solids, liquids and other SCCSs at the kinetic level.

The 2D SCDP has become particularly favorable in recent laboratory experiments, probably because complications due to the
ion-wake effect \cite{Nambu1995,Shukla1996} can be avoided in 2D systems, and thus the interaction between dust particles can be
well approximated by the Debye-H\"{u}ckel/Yukawa potential \cite{Konopka2000,Lemons2000},
$\phi(r)=(Q^2/r)\exp{(-r/\lambda_{D})}$, where $Q$ is the charge on each particle, $r$ is the inter-particle distance, and
$\lambda_D$ is the screening length. Such a system can be fully characterized in equilibrium by the screening parameter,
$\kappa=a/\lambda_{D}$, and the Coulomb coupling strength, $\Gamma=Q^2/(ak_B T)$, where $k_B T$ is the temperature of the system
and $a$ is the Wigner-Seitz radius, given by $a=(\pi n)^{-1/2}$ with $n$ being the areal density of particles. Of great interest
in 2D dusty plasmas are their collective and dynamical properties, such as the longitudinal and transverse wave modes, which
have been studied extensively over the past decade in experiments
\cite{Homann1997,Homann1998,Nunomura2002,Zhdanov2003,Nunomura2005,Nosenko2006}, theories
\cite{Dubin2000,Wang2001,Murillo2003,Kalman2004,Piel2006,Sullivan2006,Donko2008,Hou2009}, and numerical simulations
\cite{Kalman2004,Sullivan2006,Donko2008,Hou2009} (see, in particular, Ref.\ \cite{Donko2008} for a review of recent
developments).

In particular, there has been a growing interest recently in studying the collective dynamics of SCDPs in an external magnetic
field \cite{Uchida2004,Jiang2007,Jiang2007pop,Sun2007,Farokhi2009,Pilch}. However, most of those studies were theoretical. For
example, Uchida \emph{et al.}\cite{Uchida2004} derived a wave dispersion relation for a 2D Yukawa lattice in a perpendicular
magnetic field based on the harmonic approximation (HA) and compared it with the results obtained through a Molecular Dynamics
(MD) simulation in a perfect triangular lattice. Jiang \emph{et al.}\cite{Jiang2007} studied the propagation of Mach waves in a
2D magnetized Yukawa system based on the random-phase approximation (RPA), where the effect of strong coupling was neglected.
That work was subsequently extended to include the effect of strong coupling \cite{Jiang2007pop} by adopting the quasilocalized
charge approximation (QLCA) \cite{Golden1993,Kalman2004,Donko2008}. Moreover, Sun \emph{et al.} \cite{Sun2007} examined the
influence of an external magnetic field on a dust lattice wave in a plasma crystal. Very recently, Farokhi \emph{et al.}
\cite{Farokhi2009} derived analytically a wave dispersion relation in a 2D hexagonal Yukawa lattice in an external magnetic
field with arbitrary direction.

Although many theoretical proposals have arisen from the above listed studies
\cite{Uchida2004,Jiang2007,Jiang2007pop,Sun2007,Farokhi2009} and experimentlists have also attempted to study the dynamics of
magnetized dusty plasmas \cite{Pilch}, exploring these proposals in the laboratory does not seem to be feasible yet due to many
technical constraints \cite{Uchida2004,Jiang2007,Jiang2007pop,Sun2007,Farokhi2009,Pilch}. In this situation, it is desirable to
perform numerical experiments that can validate the existing theories and that can serve as a guide for future laboratory
experiments. Other than the above-mentioned MD simulation by Uchida \emph{et al.}\cite{Uchida2004}, we have found only two
related MD simulations in open literature: one studying static structures of 2D magnetized dusty plasmas by Liu \emph{et
al.}\cite{Liu2005}, and the other studying collective modes in a Coulomb system (2D strongly correlated electrons) by
Ranganathan and Johnson \cite{Ranganathan2002,Ranganathan2005}. Given that, to the best of our knowledge, there are no computer
simulations studying collective modes in magnetized Yukawa systems that cover both the liquid and solid states, we perform the
numerical experiments here to fill the gap.

First, we obtain thermally excited phonon spectra from direct simulations of 2D Yukawa systems in a perpendicular magnetic
field. The resulting dispersion relations are then compared with those derived in the above mentioned analytical theories for
different values of the screening parameter and the Coulomb coupling strength. The rest of the paper is organized as follows. In
Sec.\ II, we provide details of our simulations of 2D Yukawa solids and liquids with perpendicular magnetic field. In Sec.\ III,
analytical formulas are given for dispersion relations in the RPA, QLCA and HA, and compared with simulations. A discussion of
the results is given in Sec.\ IV, followed by a brief conclusion in Sec.\ V.

\section{Simulation}

\subsection{Algorithm}

Our simulation is performed by using the Brownian Dynamics (BD) method \cite{Allen1989, BDM2009}, in which we assume that
numerous and frequent collisions with the neutral-gas molecules give rise to a Brownian motion of each dust particle in the
plasma. These collisions give rise to the well-known Epstein drag, which is quantified by the friction coefficient, or damping
rate $\gamma$. We therefore need to introduce one additional parameter, the damping coefficient $\gamma/\omega_{pd}$, where
$\omega_{pd} = \left[2Q^{2}/(m a^{3})\right]^{1/2}$ is the nominal dusty plasma frequency, enabling us to fully characterize the
system \cite{Fortov2003}.

In our work, $N=4000$ particles are simulated in a square with periodic boundary conditions. Charged dust particles interact
with each other via pairwise Yukawa potentials, while undergoing Brownian motions. The simulation follows the trajectory of each
dust particle by numerically integrating and solving the corresponding Langevin equation and its integral
\cite{BDM2009,Lemons1999,Lemons2002},
\begin{eqnarray}
\frac{d}{dt}\mathbf{v}&=&-\gamma \mathbf{v}+\frac{1}{m}\mathbf{F}+\frac{1}{mc}Q\mathbf{\mathbf{v}\times\mathbf{B}}+\mathbf{A}(t)
,
\nonumber \\
\frac{d}{dt}\mathbf{r}&=&\mathbf{v}  \label{eqlangevin},
\end{eqnarray}
where, as usual, $\mathbf{v}$ and $\mathbf{r}$ are the velocity and position of a dust particle, respectively, and $\mathbf{F}$
is the systematic (deterministic) force coming from interactions with all other dust particles in the system. Furthermore,
$\mathbf{B}=\{0,0,B \}$ is the external magnetic field, $c$ is the speed of light in vacuum, and $\mathbf{A}(t)$ is Brownian
acceleration, respectively. Note that the friction coefficient $\gamma$ and the Brownian acceleration $\mathbf{A}(t)$ represent
complementing effects of the same sub-scale phenomenon: numerous collisions of a Brownian particle with the neutral-gas
molecules in the system. While the former represents an average effect of these collisions, the latter represents fluctuations
due to the discreteness of collisions and is generally assumed to be well represented by delta-correlated Gaussian white noise.
The friction coefficient and the Brownian acceleration are both related to the medium temperature through a
fluctuation-dissipation theorem. We adopt here a Gear-like Predictor-Corrector method for our BD simulations, which was used
successfully in simulating the shock wave propagation \cite{Jiang2006,Hou2008pop}, the heat conduction \cite{SCCS2008}, and
diffusion processes in SCDPs \cite{Hou2009PRL}. Detailed algorithms and procedures for performing these integrations can be
found in Ref.\ \cite{Hou2008pop,BDM2009}. The advantages of using the BD method are as follows. First, it brings the simulation
closer to real dusty plasma experiments by taking into account both the Brownian motion and the damping effect in a
self-consistent manner. Second, it simplifies the simulation as no external thermostat is needed.

Initially, charged dust particles are randomly placed in the square. The system comes to an equilibrium after a period
proportional to $1/\gamma$, typically around $10/\gamma$, except under conditions with extremely strong coupling and large
magnetic field, in which case relaxation of the system becomes extremely slow. One would expect a faster relaxation of the
system (and therefore a faster simulation) at higher damping rates. However, we choose here $\gamma=10^{-2}\times\omega_{pd}$ to
minimize the damping effect on collective modes, thus allowing us to concentrate on the effects of the magnetic field. We note
that the effect of neutral-gas damping on collective modes in unmagnetized SDCPs has been discussed in detail elsewhere
\cite{Hou2009}.

\subsection{Calculation of the wave spectra}

Wave spectra are determined from the current-current correlation functions. We follow the definition given in Ref.\
\cite{Boon1980}, in which the current density of the system is
\begin{equation}
\mathbf{j}(\mathbf{r},t)=\frac{1}{\sqrt{N}}\sum^{N}_{i=1}\mathbf{v}_{i}(t)\delta[\mathbf{r} -\mathbf{r}_{i}(t)]
\label{Eq_current_density_t},
\end{equation}
while its Fourier transform is given by
\begin{equation}
j_{\alpha}(\mathbf{q},t)=\frac{1}{\sqrt{N}}\sum^{N}_{i=1}v_{i\alpha}(t) e^{i\mathbf{q}\cdot\mathbf{r}_{i}(t)},
\label{Eq_current_density_kt}
\end{equation}
where $\mathbf{r}_{i}(t)$ and $\mathbf{v}_{i}(t)$ are the position and velocity of the $i$th particle at time $t$, respectively,
and
 the index $\alpha$ labels Cartesian components $x$ and $y$. Assuming that waves propagate along the $x$ direction, i.e.,
$\mathbf{q}=\{q,0\}$, one obtains the current-current correlation functions
\begin{eqnarray}
J_{\parallel}(q,t)=\langle j^{*}_{x}({q},t)j_{x}({q},0) \rangle, \nonumber \\
J_{\perp}(q,t)=\langle j^{*}_{y}({q},t)j_{y}({q},0) \rangle, \nonumber
\end{eqnarray}
for the directions parallel and perpendicular to the wave propagation direction, respectively. One recalls that, without the
magnetic field, these functions correspond to the longitudinal and transverse modes \cite{Boon1980,Hou2009}, respectively. Here
the superscript $^*$ denotes the complex conjugate while the angular bracket denotes the ensemble average. The wave spectra are
then given by the Fourier transforms of the corresponding current-current correlations \cite{Boon1980},
\begin{eqnarray}
\mathcal{J}_{\parallel}(q,\omega)=\int^{+\infty}_{-\infty}{dt\ e^{i\omega t}J_{\parallel}(q,t)}, \nonumber \\
\mathcal{J}_{\perp}(q,\omega)=\int^{+\infty}_{-\infty}{dt\ e^{i\omega t}J_{\perp}(q,t)} . \label{Eq_spectra}
\end{eqnarray}
In simulations, Eqs.\ (\ref{Eq_spectra}) are evaluated by using a discrete Fourier transform in a period of
$1310\times\omega^{-1}_{pd}$.

\section{Theories}

\subsection{General formula}
A general expression for the dispersion relation, $\omega(\mathbf{q})$, in a magnetized 2D many-body system can be determined by
solving the following equation for $\omega$ \cite{Bonsall1977,Golden1993,Uchida2004,Jiang2007pop},
\begin{equation}
\Vert \omega^{2}\mathbf{I}-\mathbf{M}(\mathbf{q})-i\mathbf{C} \Vert=0, \label{Eq_M}
\end{equation}
where $\mathbf{I}$ is the 2D unit matrix, and $i=\sqrt{-1}$. Note that the matrix
\begin{equation}
\mathbf{C}= \left [
\begin{array}{cc}
  0 & \omega\omega_{c} \\
  -\omega\omega_{c} & 0
\end{array}
\right ]
\end{equation}
accounts for the presence of an external magnetic field, with $\omega_{c}=QB/m c$ being the gyrofrequency of dust particles.
Note that, although we have neglected damping in Eq.\ (\ref{Eq_M}), it can easily be recovered by replacing $\omega^{2}$ with
$\omega(\omega+i\gamma)$. The matrix $\mathbf{M}(\mathbf{q})$ contains details of interactions between dust particles and its
form depends on the specific theory used.

For convenience, we introduce here a new parameter $\beta=\omega_{c}/\omega_{pd}$. For typical laboratory dusty plasma
parameters ($n_{0} = 1 \times 10^{8}$ cm$^{−3}$, $k_{B} T_{i} = 0.1$ eV, $k_{B} T_{e} = 3$ eV, the radius of dust particles
$r_{d} = 0.1 \mu$m, $m = 1.0 \times 10^{-15}$ g, $Q = 100$ e and $n = 100$ cm$^{−2}$), one has $\beta = 0$, $0.1$, $0.5$ and
$1$ for $B = 0$, $1.5$, $7.5$, and $15$ tesla, respectively.

\subsection{Random phase approximation (RPA)}
In the RPA, one obtains \cite{Jiang2007}
\begin{equation}
\mathbf{M}_{RPA}(\mathbf{q})=\frac{n}{m}\phi(k)\mathbf{q}\mathbf{q}, \label{Eq_M_RPA}
\end{equation}
where $\phi(q)=2\pi Q^2a/\sqrt{\kappa^2+(qa)^{2}}$ is the 2D Fourier transform of the Yukawa potential \cite{Jiang2007}.  The
dispersion relation in the RPA can then be obtained by substituting Eq.\ (\ref{Eq_M_RPA}) into Eq.\ (\ref{Eq_M})
\cite{Jiang2007} as
\begin{equation}
\omega^2(q)=\omega^2_{c}+\omega^2_{0}(q), \label{Eq_wk_RPA}
\end{equation}
where $\omega^{2}_{0}(q)=\omega^{2}_{pd}q^{2}a^{2}/\sqrt{(qa)^{2}+\kappa^2}$. It is clear that the mode given in Eq.\
(\ref{Eq_wk_RPA}) is a hybrid cyclotron-plasmon mode in 2D resulting from a superposition of the cyclotron oscillations due to
the magnetic field and a 2D dust-acoustic wave. It corresponds to the so-called magnetoplasmon mode in pure Coulomb systems
\cite{Chiu1974,Golden1993}, and is reduced to the usual 2D dust-acoustic mode when $B\rightarrow 0$ \cite{Shukla2002}. More
details of the derivation above can be found in Ref.\ \cite{Jiang2007}.

\subsection{Quasi-localized charge approximation (QLCA)}
\begin{figure}[htp]
\centering
\includegraphics[trim=15mm 140mm 12mm 35mm,clip, width=0.8\textwidth]{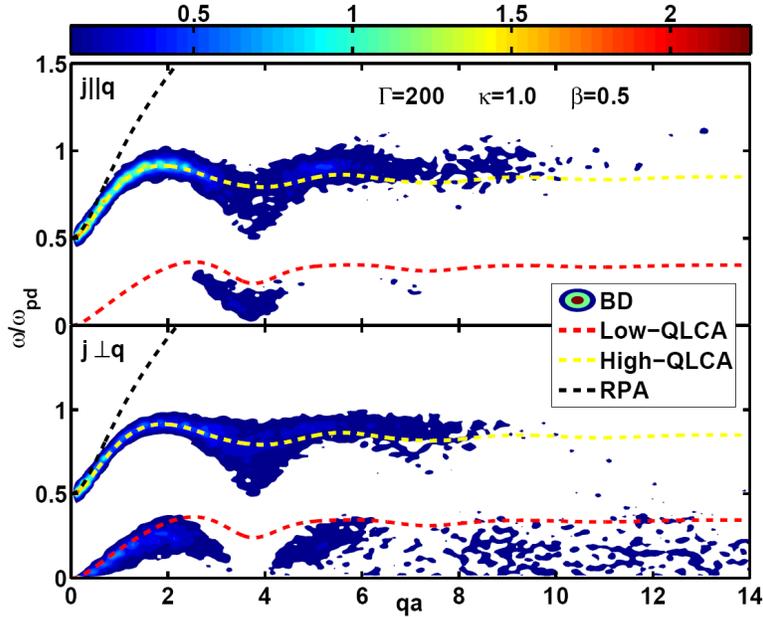}
\caption{(Color online) (Color bar in arbitrary units) The wave spectra in both the $\mathbf{j}\parallel\mathbf{q}$ and
$\mathbf{j}\perp\mathbf{q}$ directions with $\beta=0.5$, $\kappa=1$ and $\Gamma=200$. Dashed lines are the dispersion relations determined by the RPA and QLCA.}
\label{fig_B05_K1G200_0}
\end{figure}

\begin{figure}[htp]
\centering
\includegraphics[trim=0mm 0mm 12mm 0mm,clip, width=0.7\textwidth]{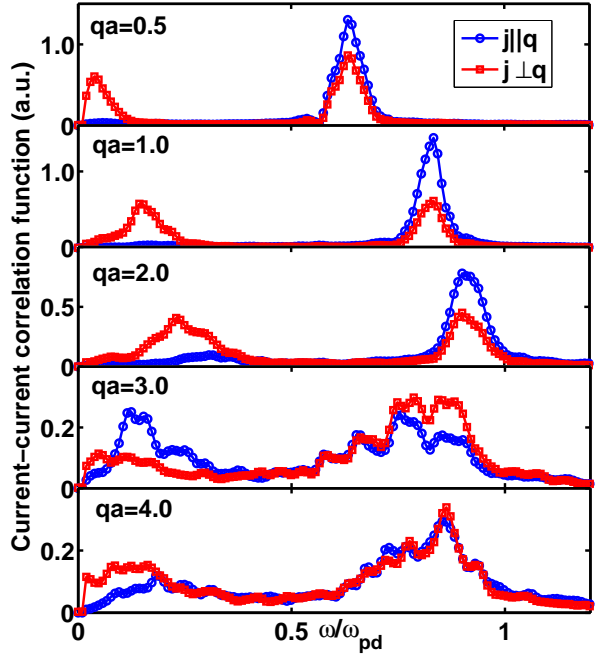}
\caption{(Color online) Profiles of the current-current correlation function in both the $\mathbf{j}\parallel\mathbf{q}$ (solid lines with
filled circles) and $\mathbf{j}\perp\mathbf{q}$ (solid lines with filled squares) directions at several fixed wavenumbers with
$\beta=0.5$, $\kappa=1$ and $\Gamma=200$.} \label{fig_B05_K1G200_a}
\end{figure}

\begin{figure}[htp]
\centering
\includegraphics[trim=15mm 140mm 12mm 35mm,clip, width=0.8\textwidth]{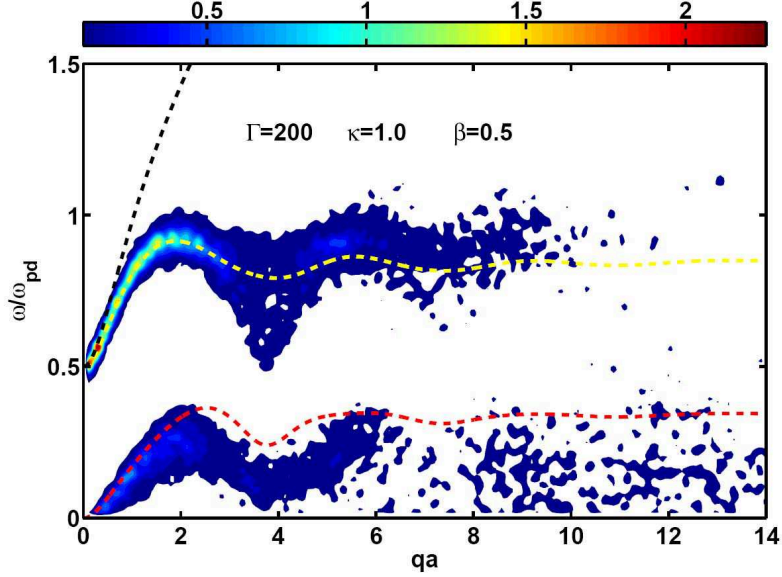}
\caption{(Color online) (Color bar in arbitrary units) The total wave spectra (sum of the $\mathbf{j}\parallel\mathbf{q}$ and
$\mathbf{j}\perp\mathbf{q}$ directions) for $\beta=0.5$, $\kappa=1$ and $\Gamma=200$. Dashed lines are the dispersion relations
determined by the RPA and the QLCA.} \label{fig_B05_K1G200}
\end{figure}

In the QLCA, one has \cite{Golden1993,Kalman2004,Donko2008,Jiang2007pop,Golden2000,Rosenberg1997,Kalman2000}
\begin{equation}
\mathbf{M}_{QLCA}(\mathbf{q})=\frac{n}{m}\phi(k)\mathbf{q}\mathbf{q} + \mathbf{D}(\mathbf{q}). \label{Eq_M_QLCA}
\end{equation}
The second term, $\mathbf{D}({\mathbf{q}})\equiv [D_{\mu\nu}(\mathbf{q})]$ (with $\mu$ and $\nu=x,y$), is the $2\times 2$
dynamical matrix of the QLCA \cite{Kalman2004,Jiang2007pop,Donko2008,Golden2000,Rosenberg1997,Kalman2000}, which takes into
account the short-range inter-particle correlations that are missing in the RPA. By neglecting this term, we immediately recover
the RPA result given in Eq.\ (\ref{Eq_M_RPA}).

For the QLCA \cite{Jiang2007pop}, the wave dispersion relation has two branches: the \emph{high}-frequency branch
$\omega_{+}(\mathbf{q})$ and the \emph{low}-frequency branch $\omega_{-}(\mathbf{q})$, which are given by
\begin{eqnarray}
\omega^2_{\pm}(q)&=&\frac{1}{2}\left [D_{L}(q)+D_{T}(q)+\omega^{2}_{0}(q)+\omega^2_{c} \right ] \nonumber \\
&\pm & \frac{1}{2}\sqrt{[D_{L}(q)+D_{T}(q)+\omega^{2}_{0}(q)+\omega^2_{c}]^2+4\omega^{2}_{c}D_{T}(q)}, \label{Eq_wk_QLCA}
\end{eqnarray}
where $D_{L}(q)$ and $D_{T}(q)$ are the projections of $\mathbf{D}(\mathbf{q})$ on the longitudinal and transverse directions,
respectively. One recognizes that the high-frequency branch is actually the magnetoplasmon mode with corrections due to
particle-particle correlations, while the low-frequency branch is identified as the magnetoshear mode, which is maintained by
the strong particle-particle correlations. When $B\rightarrow 0$, Eqs.\ (\ref{Eq_wk_QLCA}) yield the longitudinal and transverse
modes of the QLCA in unmagnetized 2D Yukawa liquids, respectively \cite{Kalman2004,Jiang2006,Hou2009}. It should be noted that
the presence of the magnetic field does not alter the general scenario of the longitudinal and transverse modes at $B=0$,
although it strongly modifies the dispersion relation. Namely, one sees that both $D_{L}(q)$ and $D_{T}(q)$ appear in both the
high- and low-frequency branches as a result of the interplay between the particle-particle correlations and the external
magnetic field. In addition, let us mention that the magnetic field has no effect on the static structure of a system in
equilibrium. Therefore, expressions for $D_{L}$ and $D_{T}$ are exactly the same as those obtained without the magnetic field
\cite{Kalman2004,Jiang2006,Donko2008,Golden2000,Hou2009,Rosenberg1997,Kalman2000}.

\subsection{Harmonic approximation (HA)}
In the HA, it is assumed that the charged dust particles form a perfect triangular lattice in 2D and are located at the points
of that lattice while performing thermal oscillations around their equilibrium positions. In this case, one needs to adopt an
another strategy to obtain the interaction matrix and the dispersion relation. One finds
\cite{Bonsall1977,Dubin2000,Donko2008,Hou2009}
\begin{equation}
\mathbf{M}_{HA}(\mathbf{q})=\frac{1}{m}\sum_{i}\frac{\partial^{2}\phi}{\partial\mathbf{r}_{i}
\partial\mathbf{r}_{i}}[1-e^{i\mathbf{q\cdot}\mathbf{r}_{i}}]\equiv [M_{\mu\nu}(\mathbf{q})],
\label{Eq_M_HA}
\end{equation}
where again $\mu$ and $\nu=x,y$, and the summation over $i$ includes all points on the triangular lattice
\cite{Bonsall1977,Dubin2000,Donko2008,Hou2009}.

The resulting dispersion relations correspond to the so-called magnetophonon modes \cite{Bonsall1977}, and are quite similar to
those of the QLCA. There are two branches, given by
\begin{eqnarray}
\omega^2_{\pm}(q,\theta)&=&\frac{1}{2}\left [ M_{xx}+M_{yy}+\omega^2_{c} \right ] \nonumber \\
&\pm & \frac{1}{2}\sqrt{[M_{xx}+M_{yy}+\omega^2_{c}]^2+4(M^{2}_{xy}-M_{xx}M_{yy})}, \label{Eq_wk_HA}
\end{eqnarray}
where $M_{\mu\nu}$ is determined by Eq.\ (\ref{Eq_M_HA}). The wave propagation depends on the polarization angle $\theta$, and
the corresponding dispersion relations are periodic functions of $\theta$ with a period of $\pi/6$ due to the hexagonal symmetry
\cite{Bonsall1977,Dubin2000,Donko2008,Hou2009}. The angular dependence arises because the system is now anisotropic at short
wavelengths.

\begin{figure}[htp]
\centering
\includegraphics[trim=15mm 140mm 12mm 35mm,clip, width=0.8\textwidth]{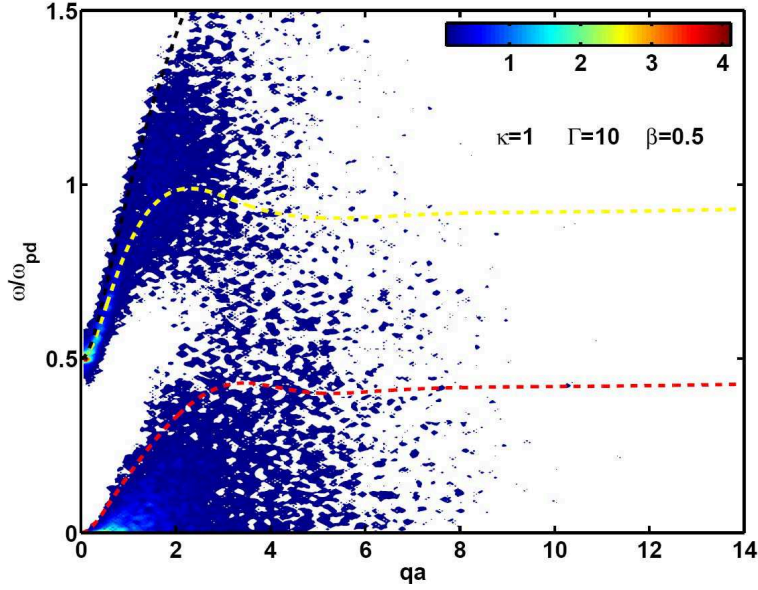}
\caption{(Color online) (Color bar in arbitrary units) The total wave spectra (sum of the $\mathbf{j}\parallel\mathbf{q}$ and
$\mathbf{j}\perp\mathbf{q}$ directions) for $\beta=0.5$, $\kappa=1$ and $\Gamma=10$. Dashed lines are the dispersion relations
determined by the RPA and the QLCA.} \label{fig_B05_K1G10}
\end{figure}

\begin{figure}[htp]
\centering
\includegraphics[trim=15mm 140mm 12mm 35mm,clip, width=0.8\textwidth]{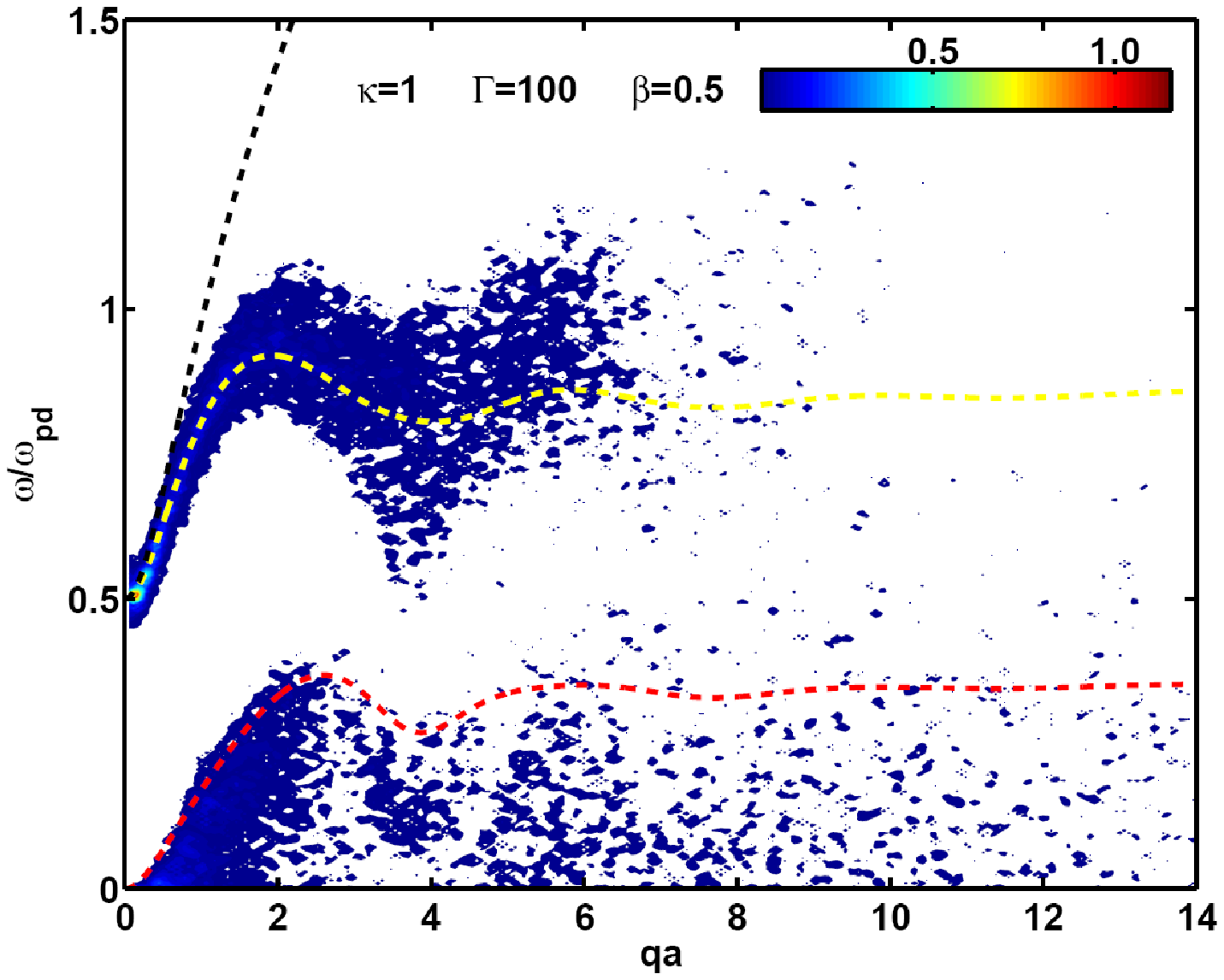}
\caption{(Color online) (Color bar in arbitrary units) The total wave spectra (sum of the $\mathbf{j}\parallel\mathbf{q}$ and
$\mathbf{j}\perp\mathbf{q}$ directions) for $\beta=0.5$, $\kappa=1$ and $\Gamma=100$. Dashed lines are the dispersion relations
determined by the RPA and the QLCA.} \label{fig_B05_K1G100}
\end{figure}

\begin{figure}[htp]
\centering
\includegraphics[trim=15mm 140mm 12mm 35mm,clip, width=0.8\textwidth]{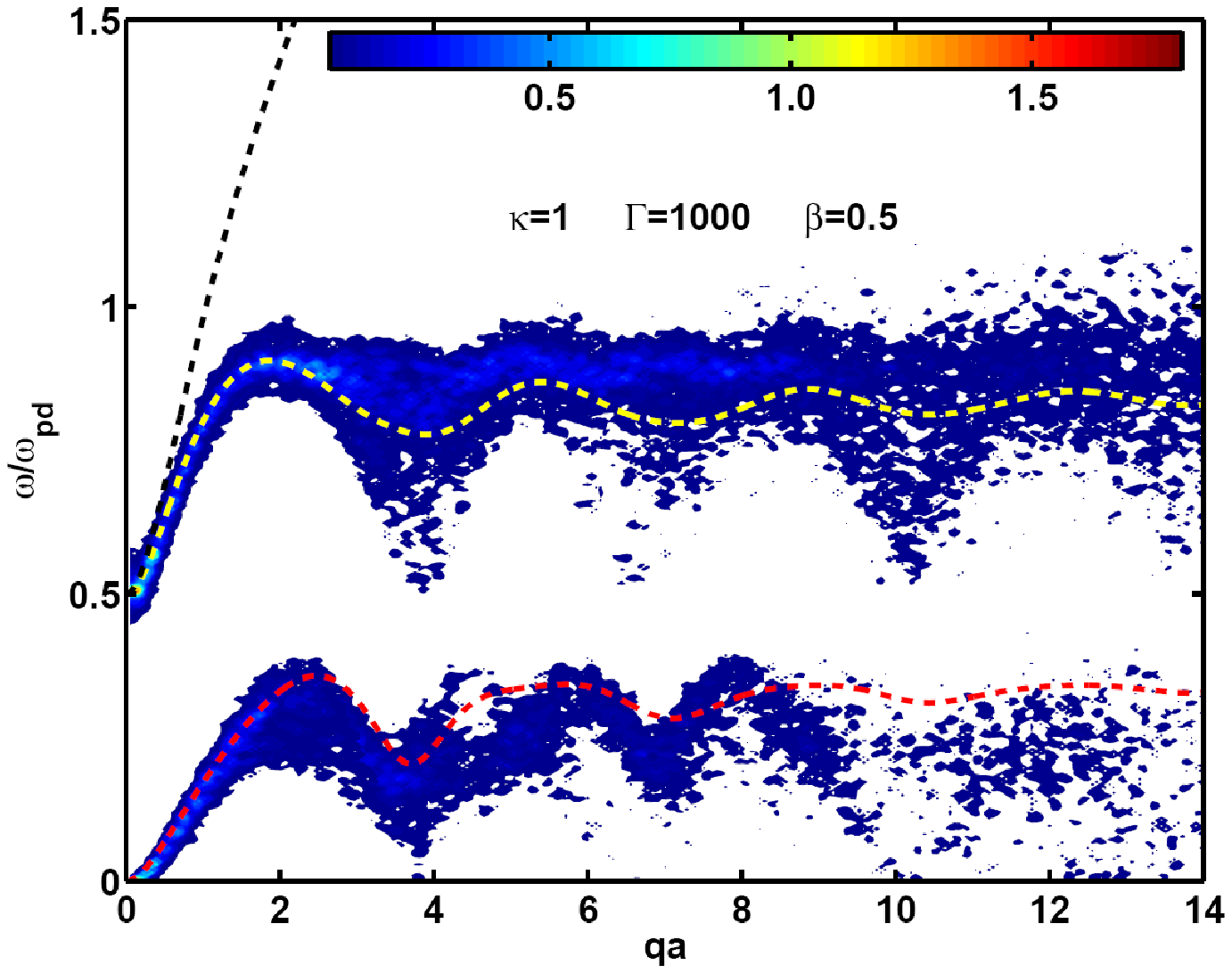}
\caption{(Color online) (Color bar in arbitrary units) The total wave spectra (sum of the $\mathbf{j}\parallel\mathbf{q}$ and
$\mathbf{j}\perp\mathbf{q}$ directions) for $\beta=0.5$, $\kappa=1$ and $\Gamma=1000$. Dashed lines are the dispersion relations
determined by the RPA and the QLCA.}
\label{fig_B05_K1G1000_EQLCA}
\end{figure}

\begin{figure}[htp]
\centering
\includegraphics[trim=15mm 140mm 12mm 35mm,clip, width=0.8\textwidth]{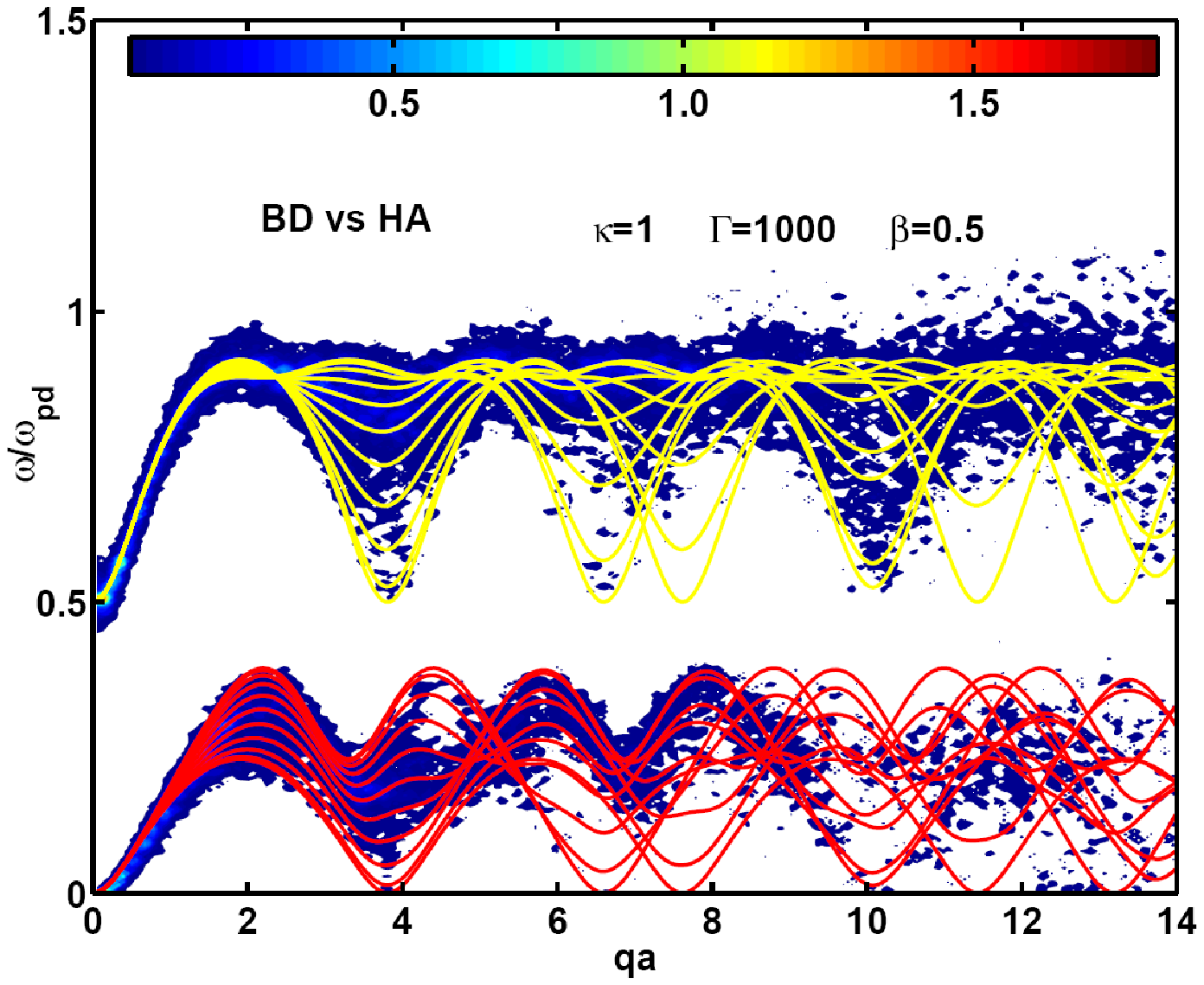}
\caption{(Color online) (Color bar in arbitrary units) The total wave spectra (sum of the $\mathbf{j}\parallel\mathbf{q}$ and
$\mathbf{j}\perp\mathbf{q}$ directions) for $\beta=0.5$, $\kappa=1$ and $\Gamma=1000$. Solid lines are the angle-dependent
dispersion relations determined by the HA, with light-gray (yellow) for the high-frequency branch and dark-gray (red) for the low-frequency branch,
respectively.} \label{fig_B05_K1G1000_HA}
\end{figure}

\section{Results and discussions}

As a typical example, we show in Fig.\ \ref{fig_B05_K1G200_0} the simulation wave spectra in both the longitudinal
($\mathbf{j}\parallel\mathbf{q}$) and transverse ($\mathbf{j}\perp\mathbf{q}$) directions for $\beta=0.5$, $\kappa=1$ and
$\Gamma=200$, along with the dispersion relations from the RPA and QLCA theories. The results clearly show that the high- and
low-frequency branches co-exist in both the longitudinal and transverse directions. The high-frequency magnetoplasmon modes
\cite{Golden1993} exhibit typical oscillations that originate from the short-range order in strongly coupled systems
\cite{Golden1993}, whereas their long wavelength limits approach the cyclotron-plasmon oscillation frequency. However, the two
branches of modes are not isotropic because the wave spectra appear to be different in the $\mathbf{j}\parallel\mathbf{q}$ and
$\mathbf{j}\perp\mathbf{q}$ directions. This is shown particularly clearly in the low-frequency branches. For example, there is
a gap at long wavelengths of the low-frequency branch in the $\mathbf{j}\parallel\mathbf{q}$ direction, and a gap around the
first trough of oscillations in the $\mathbf{j}\perp\mathbf{q}$ direction. This indicates that coupling between the longitudinal
and transverse modes strongly depends on both the wavenumber and frequency \cite{Uchida2004}. A more detailed analysis of the
frequency dependence of this coupling is shown in Fig.\ \ref{fig_B05_K1G200_a} for several wavenumbers, where one can see that
the amount of energy in the low- and high-frequency branches is redistributed between the two directions of wave propagation in
a complementary manner. For example, when analyzing long wavelengths, one finds more energy in the low-frequency branch for the
transverse direction than for the longitudinal direction, while there is more energy in the high-frequency branch for the
longitudinal direction than for the transverse direction. The roles of the transverse and longitudinal directions can be
reversed at other wavelengths, as shown in Fig.\ \ref{fig_B05_K1G200_a} for $qa=3$. We have observed in our simulations that
such an effect occurs for any value of the coupling strength, and is more pronounced for higher $\beta$s. We note that a similar
effect was observed previously in the MD simulation of a perfect triangular lattice \cite{Uchida2004}.

In the following, we shall simply add the wave spectra obtained in both directions to achieve a more complete picture of both
the high- and low-frequency branches. This is done in Fig.\ \ref{fig_B05_K1G200} using the data from Fig.\
\ref{fig_B05_K1G200_0}, where both branches now show similar oscillatory structures. One observes that the QLCA captures the
dispersion of the high-frequency wave spectra very well in a reasonably large range of wavelengths, whereas the RPA displays
noticeable discrepancy with our simulations except in the long wavelength limit. For the low-frequency branch, the agreement
between the QLCA and simulations is less satisfactory, although the QLCA seems to give the correct trend in the oscillatory
structures of the spectra.

We first examine how collective dynamics depends on the state of the system. Figures \ref{fig_B05_K1G10}, \ref{fig_B05_K1G100},
and \ref{fig_B05_K1G1000_EQLCA} show the wave spectra for a high temperature liquid with $\Gamma=10$, a cool liquid with
$\Gamma=100$, and a crystalline state with $\Gamma=1000$, respectively, where $\kappa=1$ and $\beta=0.5$ are kept fixed in all
three figures. One sees from Fig.\ \ref{fig_B05_K1G10} that, at $\Gamma=10$, both the high- and low-frequency modes are heavily
damped at short wavelengths, say $qa>1$, and there are essentially no collective modes beyond $qa=6$. One would attribute this
to Landau damping and viscous/collisional damping. The QLCA dispersion relation again agrees reasonably well with our
simulations for the high-frequency branch, in the region where the peak in the spectra is distinguishable, while the agreement
between the QLCA and our simulations is poorer for the low-frequency branch. There may be several reasons for this. On one hand,
since the low-frequency mode is always heavily damped in the simulations, the corresponding wave spectra spread out in the
$\omega-q$ plane, and it is difficult to isolate their peak positions. On the other hand, it is well known that most of the
damping effects, such as Landau damping and diffusional damping, are not included in the QLCA
\cite{Golden1993,Golden2000,Rosenberg1997,Kalman2004,Kalman2000}. This can be related to the inability of the QLCA to predict
the long wavelength cutoff in the shear mode, which is clearly seen in the low-frequency branch in Fig.\ \ref{fig_B05_K1G10}.

As expected, the damping effect becomes much weaker with increasing $\Gamma$ or, equivalently, with decreasing temperature. When
$\Gamma=100$, one observes from Fig.\ \ref{fig_B05_K1G100} that the wave spectra are largely extended in the short wavelength
direction, and so is the region of agreement between the QLCA and the simulations for the high-frequency branch. However, the
discrepancy between the QLCA and the simulations for the low-frequency branch is still significant. A substantial improvement is
observed only in Fig.\ \ref{fig_B05_K1G1000_EQLCA} at $\Gamma=1000$, for which the long wavelength cutoff in the low-frequency
branch disappears and oscillatory structures are quite distinguishable in both branches. A good agreement between the QLCA and
simulations is found in both branches, particularly in the first Brillouin zone. For shorter wavelengths, some fine structures
appear in the spectra due to the angular dependence of the wave dispersion, and the QLCA dispersion can only give a correct
tendency of the oscillations. The task of resolving full angular dependence of the wave dispersion lies beyond the capability of
the QLCA, but this issue can be tackled by the HA, as shown next.
\begin{figure}[htp]
\centering
\includegraphics[trim=15mm 140mm 12mm 35mm,clip, width=0.8\textwidth]{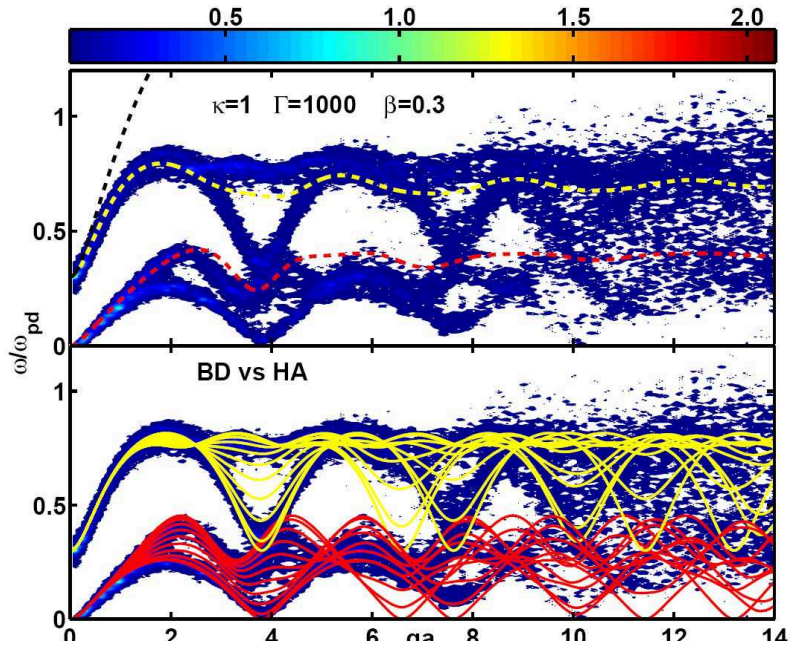}
\caption{(Color online) (Color bar in arbitrary units) The total wave spectra (sum of the $\mathbf{j}\parallel\mathbf{q}$ and
$\mathbf{j}\perp\mathbf{q}$ directions) for $\beta=0.3$, $\kappa=1$ and $\Gamma=1000$. Dashed lines are the dispersion relations
determined by the RPA and the QLCA. Solid lines are the
angle-dependent dispersion relations determined by the HA, with light-gray (yellow) for the high-frequency branch and dark-gray (red) for the low-frequency branch,
respectively.} \label{fig_B03_K1G1000}
\end{figure}

Figure \ref{fig_B05_K1G1000_HA} displays the same spectra as shown in Fig.\ \ref{fig_B05_K1G1000_EQLCA}, but comparisons are
made with the dispersion results from the HA. There are two groups of dispersions from the HA, corresponding to the high- and
low-frequency modes, respectively, and in each group ten different curves that cover the period of $\pi/6$ evenly are shown. One
sees that the structure of the wave spectra is well overlapped by the HA dispersion curves, indicating a good agreement between
the HA and our simulation. By comparing the results of the HA and those of the QLCA in Fig.\ \ref{fig_B05_K1G1000_EQLCA}, one
concludes that the QLCA gives a kind of angle-averaged dispersion.
\begin{figure}[htp]
\centering
\includegraphics[trim=15mm 140mm 12mm 35mm,clip, width=0.8\textwidth]{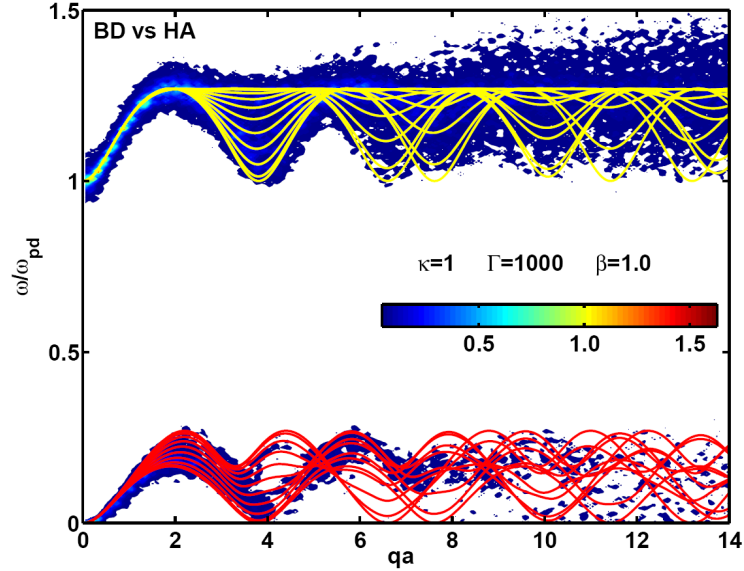}
\caption{(Color online) (Color bar in arbitrary units) The total wave spectra (sum of the $\mathbf{j}\parallel\mathbf{q}$ and
$\mathbf{j}\perp\mathbf{q}$ directions) for $\beta=1.0$, $\kappa=1$ and $\Gamma=1000$. Solid lines are the angle-dependent
dispersion relations determined by the HA, with light-gray (yellow) for the high-frequency branch and dark-gray (red) for the low-frequency branch,
respectively.} \label{fig_B10_K1G1000}
\end{figure}

\begin{figure}[htp]
\centering
\includegraphics[trim=15mm 140mm 12mm 35mm,clip, width=0.8\textwidth]{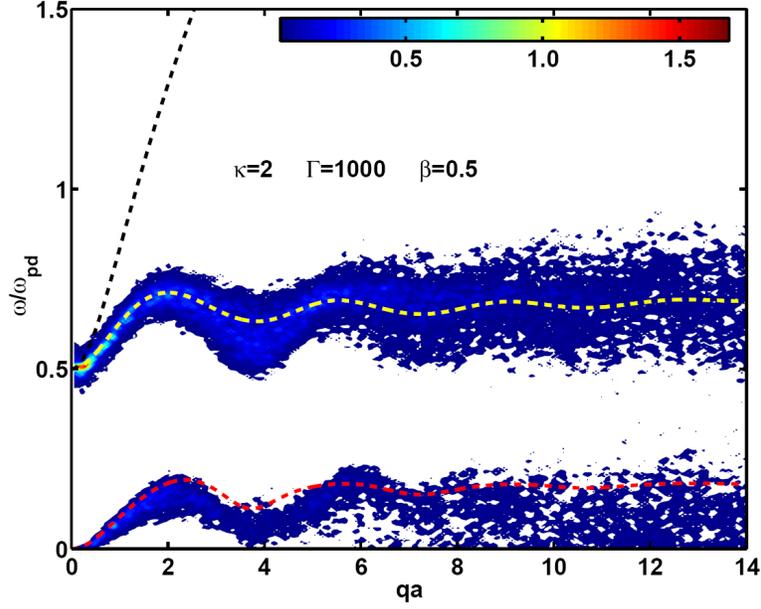}
\caption{(Color online) (Color bar in arbitrary units) The total wave spectra (sum of the $\mathbf{j}\parallel\mathbf{q}$ and
$\mathbf{j}\perp\mathbf{q}$ directions) for $\beta=0.5$, $\kappa=2$ and $\Gamma=1000$. Dashed lines are the dispersion relations
determined by the RPA and the QLCA.} \label{fig_B05_K2G1000}
\end{figure}

Next, we analyze the dependence of wave spectra on the intensity of the external magnetic field. Figure \ref{fig_B03_K1G1000}
shows wave spectra for $\beta=0.3$ (with other parameters being the same as those in Fig.\ \ref{fig_B05_K1G1000_HA}), together
with the dispersions from both the QLCA and the HA. Comparing with Fig.\ \ref{fig_B05_K1G1000_HA}, one sees from Fig.\
\ref{fig_B03_K1G1000} that the oscillations in the wave spectra are greatly enhanced in this case, and fine structures due to
the angular dependence in the wave propagation direction are enriched. As before, the agreement with the QLCA is clearly shown
for both branches, roughly within the first Brillouin zone. On the other hand, one sees that the HA captures the fine structures
in the wave spectra very well. The fact that the wave spectra in the high-frequency branch exhibit a gap in certain directions
of propagation at the wavelengths $6<qa<7$ can be rationalized as follows. Since our simulation of the crystal lattice
unavoidably has defects and contains distinct domains with different orientations, the measured wave spectra are expected to
exhibit some preference to certain angles of propagation based on the dominant crystal orientation. One can roughly estimate
such an effect by comparing the spectral distributions with the HA dispersion relations, as illustrated in Fig.\
\ref{fig_B03_K1G1000}. On the other hand, both the oscillations and fine structures in the wave spectra are suppressed for
stronger magnetic fields, as shown in Fig.\ \ref{fig_B10_K1G1000} for $\beta=1.0$, along with the dispersions from the HA. One
sees in that figure that the high-frequency branch is raised above $\omega/\omega_{pd}=\beta=1.0$ and its width is greatly
reduced, as is the width of the low-frequency branch, when the magnetic field increases.
\begin{figure}[htp]
\centering
\includegraphics[trim=15mm 140mm 12mm 35mm,clip, width=0.8\textwidth]{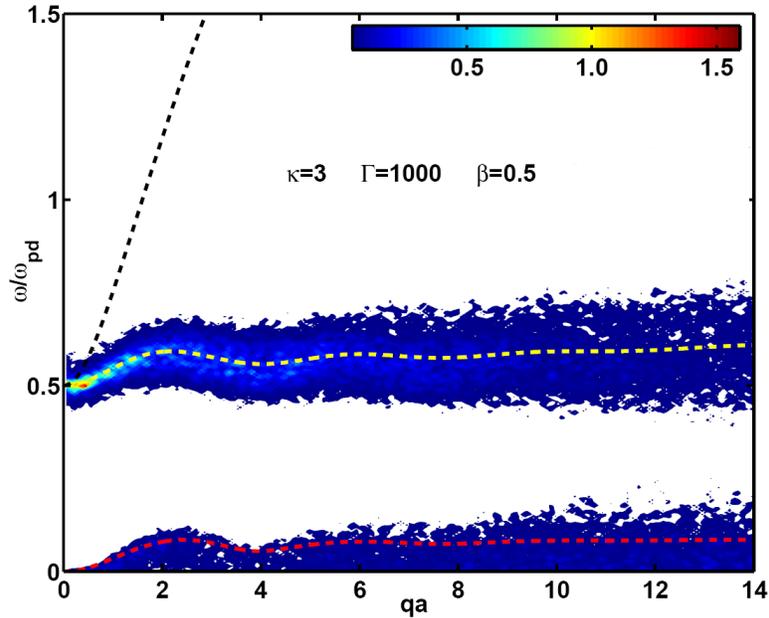}
\caption{(Color online) (Color bar in arbitrary units) The total wave spectra (sum of the $\mathbf{j}\parallel\mathbf{q}$ and
$\mathbf{j}\perp\mathbf{q}$ directions) for $\beta=0.5$, $\kappa=3$ and $\Gamma=1000$. Dashed lines are the dispersion relations
determined by the RPA and the QLCA.} \label{fig_B05_K3G1000}
\end{figure}

Finally, we examine the effect of screening on the wave spectra. Figures \ref{fig_B05_K2G1000}, \ref{fig_B05_K3G1000}, and Fig.\
\ref{fig_B05_K1G1000_EQLCA} compare the wave spectra for screening parameters $\kappa=2$, 3 and 1, respectively, with
$\Gamma=1000$ and $\beta=0.5$ kept fixed. The dispersion relations from the RPA and QLCA are also shown for comparison. It is
seen that, with increasing $\kappa$, the oscillatory structures in the spectra gradually flatten out, so that the dispersions
become almost featureless when $\kappa=3.0$. At the same time, the width of the low-frequency branch decreases continuously with
increasing $\kappa$ and, for $\kappa=3.0$, the spectra become very blurry due to heavy damping.

\section{Conclusions}

In this paper, we have used the Brownian dynamics simulation to generate wave spectra of 2D strongly coupled dusty plasmas in an
external magnetic field. We have examined how these spectra are affected by the coupling strength in such systems and by the
intensity of the magnetic field. In particular, we have compared the dispersion relations resulting from the simulation wave
spectra with those obtained from several analytical theories, including the random-phase approximation, quasi-localized charge
approximation and the harmonic approximation. Two branches of collective modes are generally observed in our simulation: the
high-frequency magnetoplasmon and the low-frequency magnetoshear modes. The RPA can only reproduce the high-frequency branch,
i.e., the magnetoplasmon mode, and it gives a good approximation only in the long wavelength limit, but remains robust for
increasing coupling strengths, even up to the crystalline state. The QLCA can reproduce both of the above modes, and it shows a
particularly good agreement with the simulation in the high-frequency branch for wide ranges of coupling strengths, screening
parameters and magnetic field magnitudes. However, the agreement between the QLCA dispersion relations and the simulation
spectra is generally poor in the low-frequency branch, particularly in a high-temperature liquid state, probably because Landau
and diffusional damping are neglected in the QLCA. The situation becomes better in a crystalline state where a good agreement
between the QLCA and the simulation is found in the low-frequency branch within the entire first Brillouin zone. However, the
QLCA cannot capture fine structures in the wave spectra due to angular-dependence of the wave propagation in the crystalline
state. In this case, it is found that the HA can give a very good description of the collective dispersions observed in the
simulation.

Our simulations show that the wave spectra exhibit profiles at fixed wavenumbers that have variable widths in the frequency
domain. These widths generally correspond to various dissipation mechanisms that give rise to damping of the collective dynamics
in strongly coupled systems. We note that we have deliberately chosen a very weak collisional damping on the neutral-gas
molecules in order to reveal the effects of the magnetic field and other parameters. While comparisons of the observed widths
cannot be made with the above analytical theories, we have found some general trends in our simulations. The strongest effect on
the frequency widths of wave spectra was found to come from the coupling strength, with the widths strongly increasing with
increasing temperature of the system. This is similar to the observations in unmagnetized SDCPs \cite{Hou2009}. With regards to
the intensity of the magnetic field, the spectra were found to be narrower at higher magnetic fields, whereas lower magnetic
fields are accompanied by broad, fine structures in the spectra reflecting a strong anisotropy in the direction of propagation.
These structures are generally very well captured by the families of dispersion relations coming from the HA. Finally, we have
found that increasing the screening parameter gives rise to narrower and flatter wave spectra, with the low-frequency branch
reduced to a heavily-damped mode akin to low-frequency noise. Clearly, more theoretical work is necessary to fully understand
the dissipation mechanisms in strongly-coupled dusty plasmas in the presence of a magnetic field.

\begin{acknowledgments} L.J.H.\ acknowledges support from the Alexander von Humboldt Foundation. The work at CAU is supported by the DFG through the SFB-TR24/A2. Z.L.M.\ acknowledges support from NSERC.
\end{acknowledgments}

\end{document}